# Missing Value Estimation Algorithms on Cluster and Representativeness Preservation of Gene Expression Microarray Data

Marie Li[1]

[1]Timpview High School, 3570 Timpview Dr., Provo UT, 84604, USA

**Abstract**

Missing values are largely inevitable in gene expression microarray studies. Data sets often have significant omissions due to individuals dropping out of experiments, errors in data collection, image corruptions, and so on. Missing data could potentially undermine the validity of research results - leading to inaccurate predictive models and misleading conclusions. Imputation - a relatively flexible, general purpose approach towards dealing with missing data - is now available in massive numbers, making it possible to handle missing data. While these estimation methods are becoming increasingly more effective in resolving the discrepancies between true and estimated values, its effect on clustering outcomes is largely disregarded.

This study seeks to reveal the vast differences in agglomerative hierarchal clustering outcomes estimation methods can construct in comparison to the precision exhibited (presented through the cophenetic correlation coefficient) in comparison to their high efficiency and effectivity in value preservation of true and imputed values (presented through the root-mean-squared error). We argue against the traditional approach towards the development of imputation methods and instead, advocate towards methods that reproduce a data set's original, natural cluster.

By using a number of advanced imputation methods, we reveal extensive differences between original and reconstructed clusters that could significantly transform the interpretations of the data as a whole.

**Introduction**

During the past decades, microarray technology has become a fundamental tool to genomics and biomedical research. Notably, gene expression profiling has allowed for countless novel developments in the study of a vast range of biological processes, such as the determination of bacterium resistance [1] and prediction of the efficacy of chemotherapy [2]. Commonly, these innovations are a result of cluster analyses, where researchers assemble groups of genes with related expression patterns to hypothesize the functional role of unknown genes and propel further research with those predictions. Additionally, statistical treatments of data are also often implemented on gene expression data with (supervised learning algorithms) [3] or without (unsupervised learning algorithms) [4] other prior information. However, almost all of these tools cannot be implemented successfully without a complete data set. Unfortunately, microarray data sets are often massive and, inevitably, frequently suffer from missing values.

Missing values in microarray experiments are generally classified into three categories: 'missing completely at random' (MCAR, where missing values occur due to technical problems unrelated to the experimental objective), 'missing at random' (MAR, where missing values occur where measurements are unreliable) , and 'missing not at random' (MNAR, where missing values are related to the experimental objective). Regardless, analyzing a data set with missing values, can lead to a reduction in the data's level of rigor, hinder downstream analyses, and can bias the inferences in experiments. [5] Two main approaches towards handling missing values are considered: omission, where samples with missing values are entirely disregarded in analysis, and imputation, where missing values are replaced with estimated values, often through statistical models. When the number of missing values are large, it is unreasonable to simple discard samples with missing values. Doing so would lead to the loss of potentially revolutionary data, introduce devastating biases and waste valuable resources. Thus, a more practical approach is to apply imputation methods to fill in missing values. Another alternative is filling missing values with some constant, such as zeroes. However, this approach fails to consider the trends and structure of the data. Hence, a more rigorous and reliable method to fill missing values is to use statistical models that capture the trends of non-missing values and reflect it upon the samples containing missing values.

Currently, there is a growing number of imputation algorithms available on the public domain that base their estimations upon predictive models such as, implementing support vector regression imputation using quadratic programming [6] and Gaussian mixture clustering imputation using expectation-maximum algorithms [7]. Frequently, imputation models are simply built upon a single goal, that is, producing values that are similar to original values. However, this approach overlooks the changes that occur in clustering outcomes, oddly, a factor that often undergoes notable changes across estimation outcomes. We argue that a greater emphasis needs to be placed upon clustering outcomes during the evaluation of estimation methods.

In this study, we systematically evaluate imputation methods based upon their ability to preserve natural cluster groups and the representativeness of gene expression microarrays: multiple imputation with a bootstrap expectation-maximum algorithm (EMB) [8], multiple imputation by chained equations (MICE) [9], local least squares (LLS) [11], single value decomposition (SVD) [12], and Bayesian Principal-Component Analysis (bPCA) [10].

**Table 1: Data Sets**

| Name  | NV        | G      | S  | MC     | MV   | NNV       | SD        |
|-------|-----------|--------|----|--------|------|-----------|-----------|
| 54536 | 472,310   | 47,231 | 10 | 47,231 | 0%   | 472,310   | 97,398.58 |
| 6248  | 2,309,331 | 45,281 | 51 | 45,281 | 0%   | 2,309,331 | 2,122.023 |
| 5473  | 2,309,136 | 48,107 | 48 | 47,323 | 1.6% | 2,271,504 | 1.063843  |
| 2490  | 245,113   | 22,283 | 11 | 22,283 | 0%   | 245,113   | 1.447674  |

NV is the number of values in the data set, G is the number of genes (number of columns in the data set), S is the number of individuals (number of rows in the data set), MC is the number of genes without one or more missing values, MV is the original percentage of missing values in the data set, NNV is the number of values in the data set after genes (rows) with one or more missing values have been removed, SD is the standard deviation of the data set after genes (rows) with one or more missing values have been removed.

**Results**

We tested imputation algorithms on four different, real gene expression microarray datasets (see Table 1), allowing this comparison to better reflect the algorithms' performance in practice. Each data set was preprocessed, filtered and reconstructed through replacing $n\%$ of the total values with missing values, namely 0.1%, 0.5%, 1%, 5%, 10%, and 15%. Five imputation algorithms were then applied to each reconstructed data set, that is, EMB, MICE, LLS, SVD, and bPCA. This procedure was iterated 20 times per data sets and rate of missingness, generating a total of 2880 unique data sets for analysis. The root mean squared error (RMSE) was used to evaluate the degree of variation between the original and imputed data values. Then, hierarchical agglomerative clustering methods were applied to each data set, generating 2880 dendrograms for analysis. A simple cophenetic correlation coefficient was then calculated to evaluate the degree of variation between the pairwise distances between the original and imputed models. The results were summarized over degrees of missingness and method of imputation (see Figure 1).

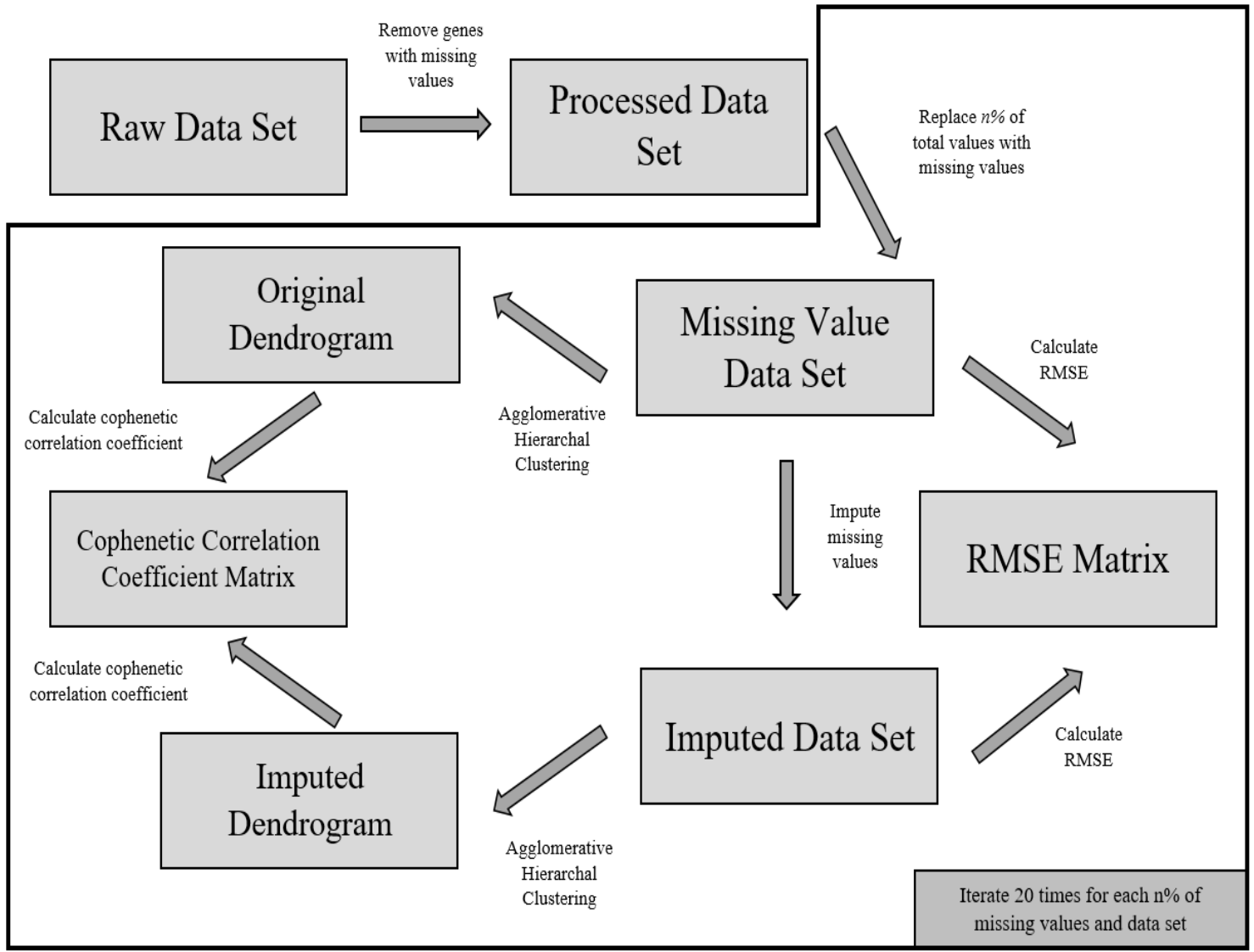

**Figure 1:** Raw data sets are first processed (by removing genes with missing values and other preprocessing procedures – see Methods). Then *n%* (0.1%, 0.5%, 1%, 5%, 10%, 15%) of total values are replaced with missing values. Data is imputed with all five different imputation methods (EMB, MICE, LLS, SVD, and bPCA) per percentage of missing values and a root-mean-squared-error is calculated. Then, each imputed data set was clustered through agglomerative hierarchal clustering and a cophenetic correlation coefficient was calculated to evaluate its performance. This process is iterated 20 times per data set and percentage of missing values.

**Discussion on Representativeness Preservation**

Individual data set imputation accuracies are summarized in Figure 2. Imputation accuracies of large and small data sets are summarized in Figure 4 (a summary of clustering preservation is also depicted in Figure 4). An overall summary of imputation accuracy is shown in Figure 5.

In Figure 2, a general trend of RMSE increasing (less imputation accuracy) as imputation percentages increases is demonstrated with few exceptions: a large decrease in RMSE at 10 - 15% in data set 2490 and a slight decrease of RMSE at 0.5 - 1% in data set 6248. The large decrease of RMSE at 10 - 15% in data set 2490 is reflected in Figure 3 for small data sets due to RMSE's natural sensitivity to outliers. [25], [26] Data sets 54536 and 2490 also have a greater range of RMSE values (54536: 2.25 – 3.25; 2490: 2.5 – 4.25) when compared to the range of RMSE values in data sets 5473 (2.4 – 3.1) and 6248 (2.45 – 2.65). The RMSE values of data sets 2490 and 5473 are much tighter between imputation algorithms than those of data sets 6248 and 54536.

Across all data sets, he bPCA and SVD imputation methods also exhibit slightly lower RMSE values in smaller data sets than other imputation methods, which represents a better fit of data, most significantly in data set 54536. MICE and EM imputation methods have the highest RMSE values, representing a lesser fit of data.

**Discussion on Cluster Preservation**

Individual data set clustering performances are summarized in Figure 3. Clustering performances of large and small data sets are summarized in Figure 4 (a summary of imputation accuracies is also depicted in Figure 4). An overall summary of clustering performances is shown in Figure 5.

The cophenetic correlation coefficients are significantly less strong and rigid – both its range of values and confidence intervals are massive. In data sets 54536, 6248, and 5473, the bPCA and SVD imputation methods, similar to its RMSE values, exhibit lower cophenetic correlation coefficients (0.725), which represents a lesser fit of data. In data set 2490, methods bPCA and SVD primarily occur at a higher coefficient: 0.9. Comparatively, in data sets 54536 and 5473, the methods EM, LLS, and MICE produce clustering results that are consistently high, representing greater similarity between the original and reconstructed dendrograms. Data set 2490 uniquely demonstrates 'messy' data, most likely due to the individual differences between the data sets themselves. These trends are reflected similarly over all data sets (Figure 5) and between large and small data sets (Figure 4). Overall, the EM, LLS, and MICE imputation methods rise with the most similar clustering outcomes, as opposed to the bPCA and SVD imputation methods, who exhibit lower correlation coefficients, representing less fit between dendrograms.

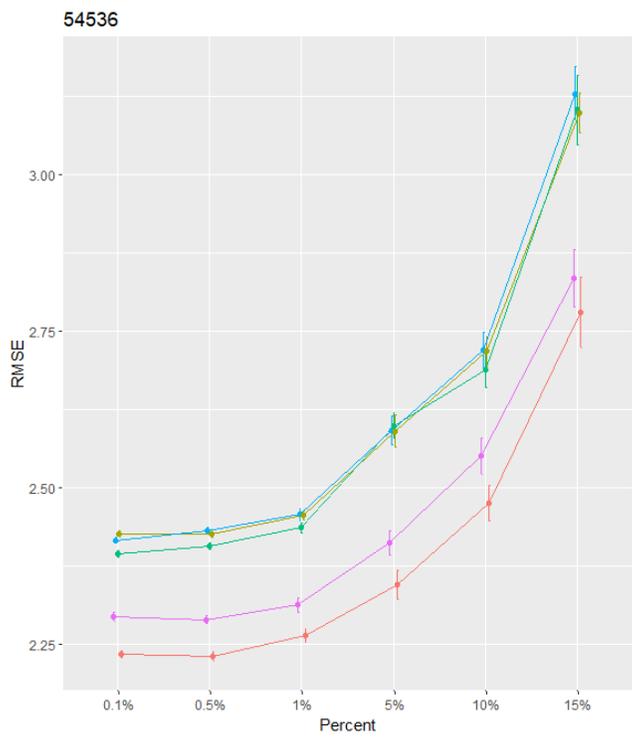
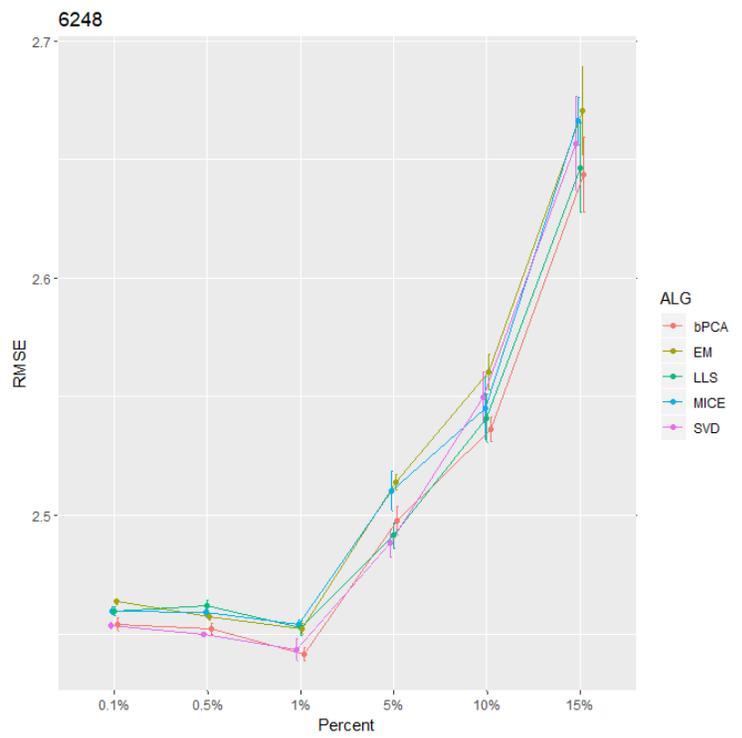
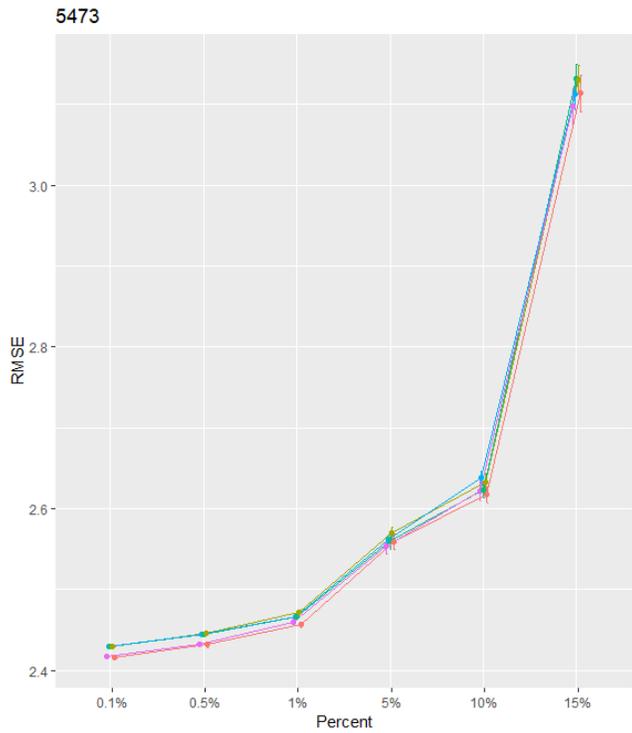
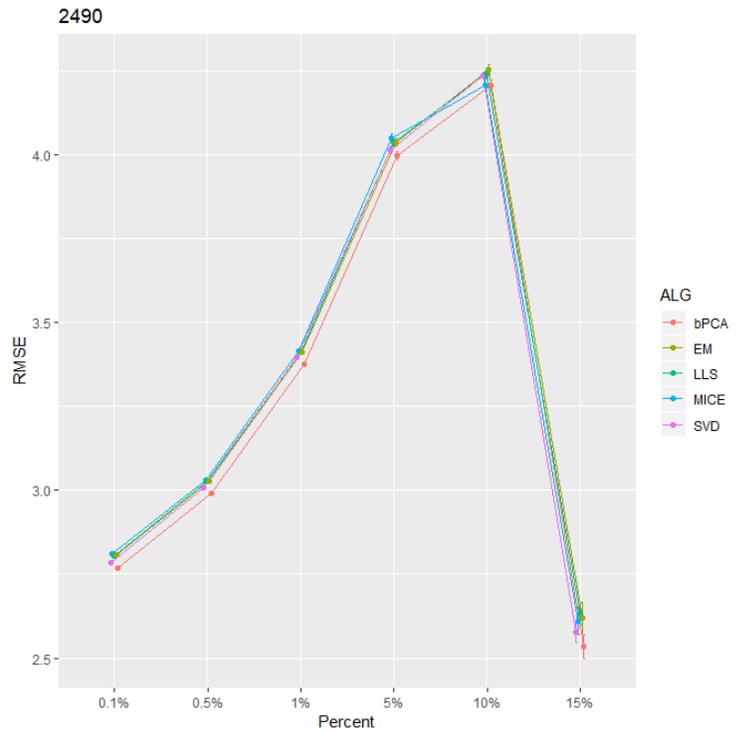

**Figure 2: Preservation of Representativeness.** The capability of each imputation method over different percentages of missing values to reproduce original data values is presented. The mean of all calculated root-mean-squared errors over 20 iterations is used to display overall imputation accuracy. An error bar calculated from a 95% confidence interval is also depicted

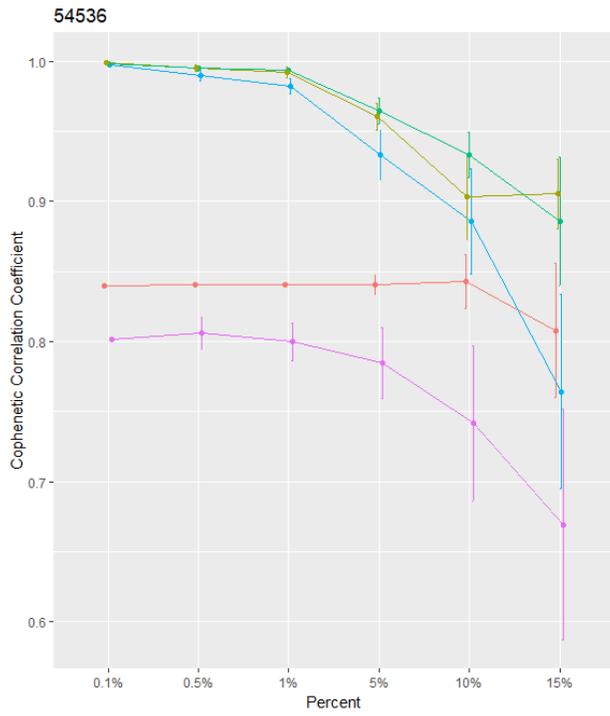 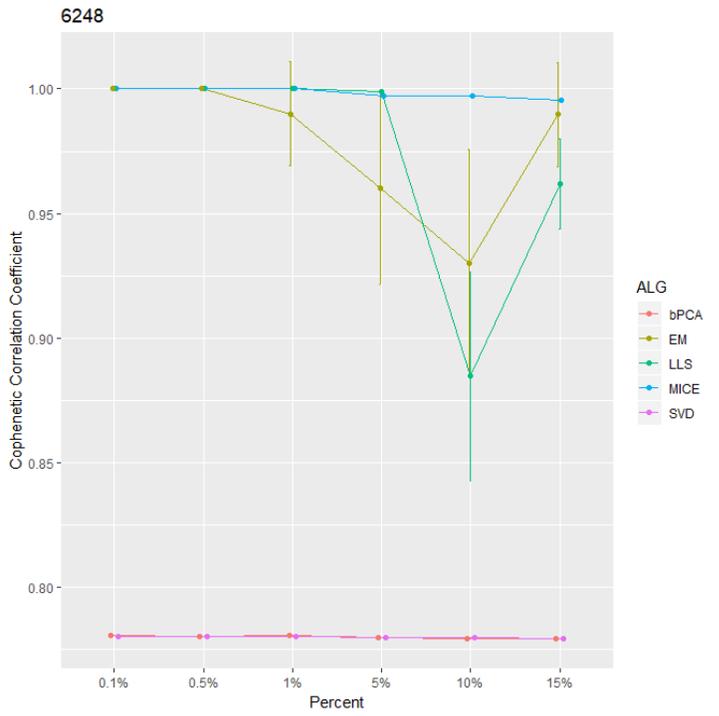
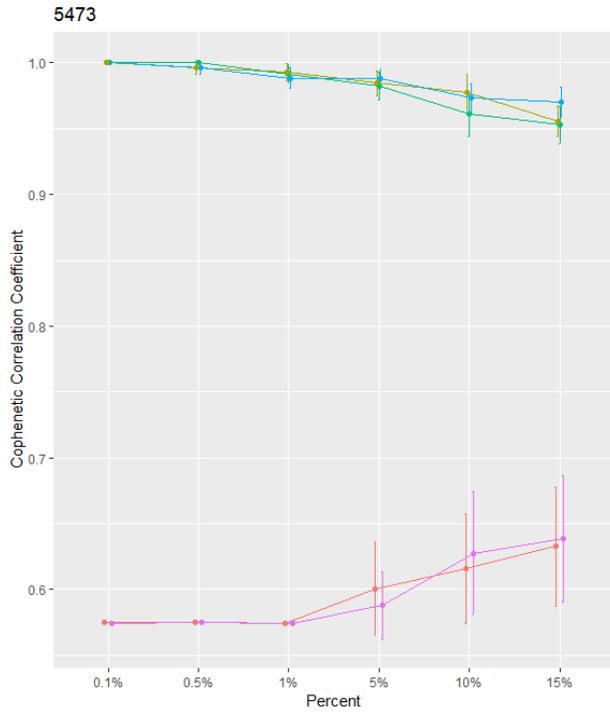 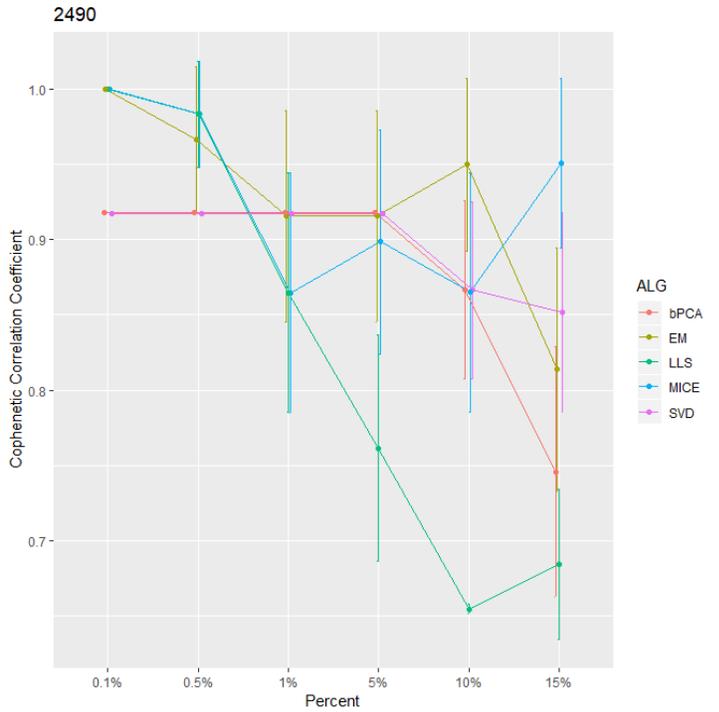

**Figure 3: Preservation of Agglomerative Hierarchal Clustering Outcomes.** The capability of each imputation method over different percentages of missing values to reproduce original clustering outcomes is presented. The mean of all calculated cophenetic correlation coefficients over 20 iterations is used to display overall clustering accuracy. An error bar calculated from a 95% confidence interval is also depicted.

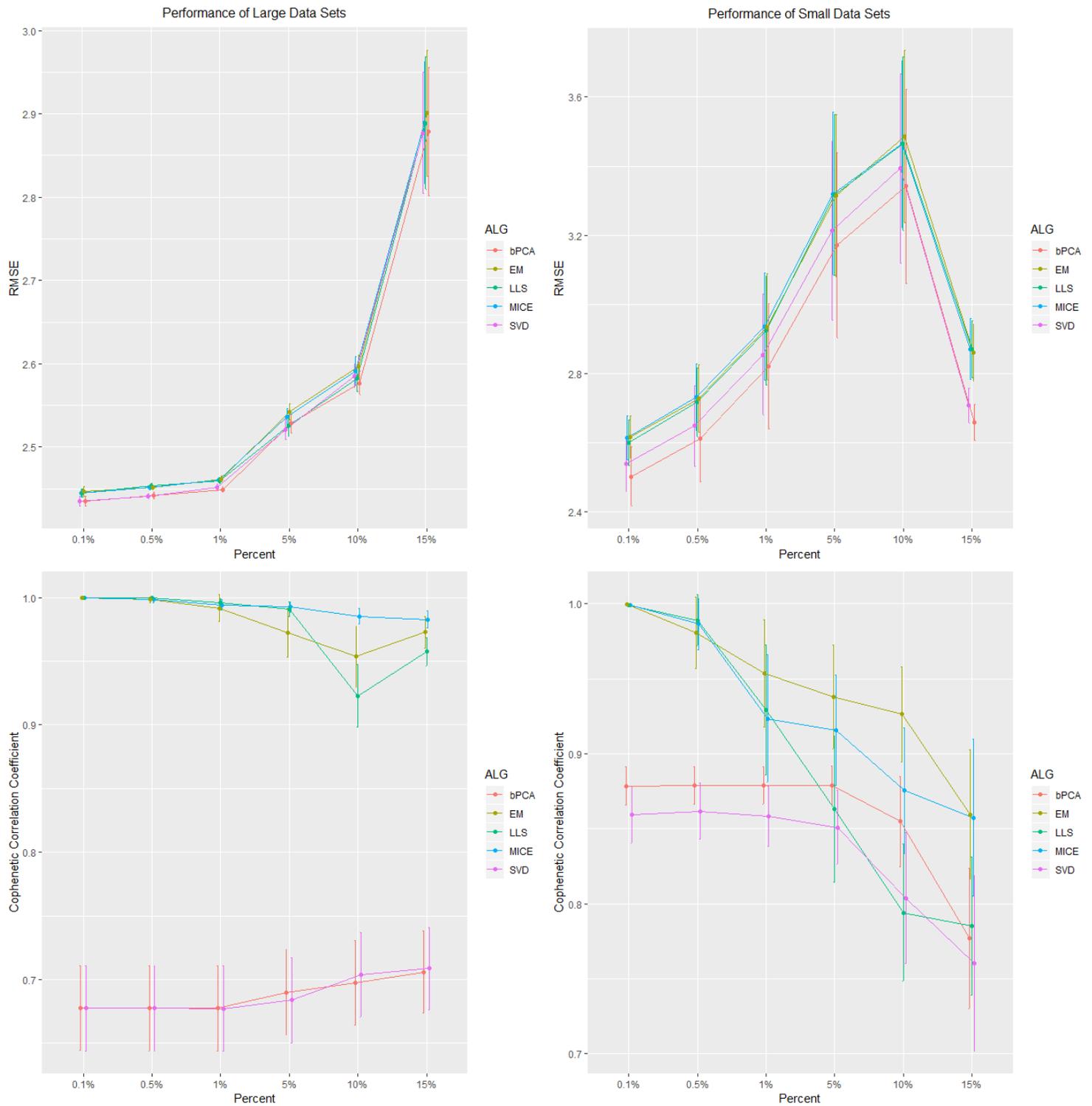

**Figure 4: Comparison of Imputation Method Performance between Large and Small Data Sets.** The capability of each imputation method over different percentages of missing values to reproduce original clustering outcomes and original values is presented. The right column of diagrams represents a summary of the performance of imputation methods on small data sets (number of values < 1,000,000) whereas the left column of diagrams summarizes the performance of imputation methods on large data sets (number of values > 1,000,000). An error bar calculated from a 95% confidence interval is also depicted

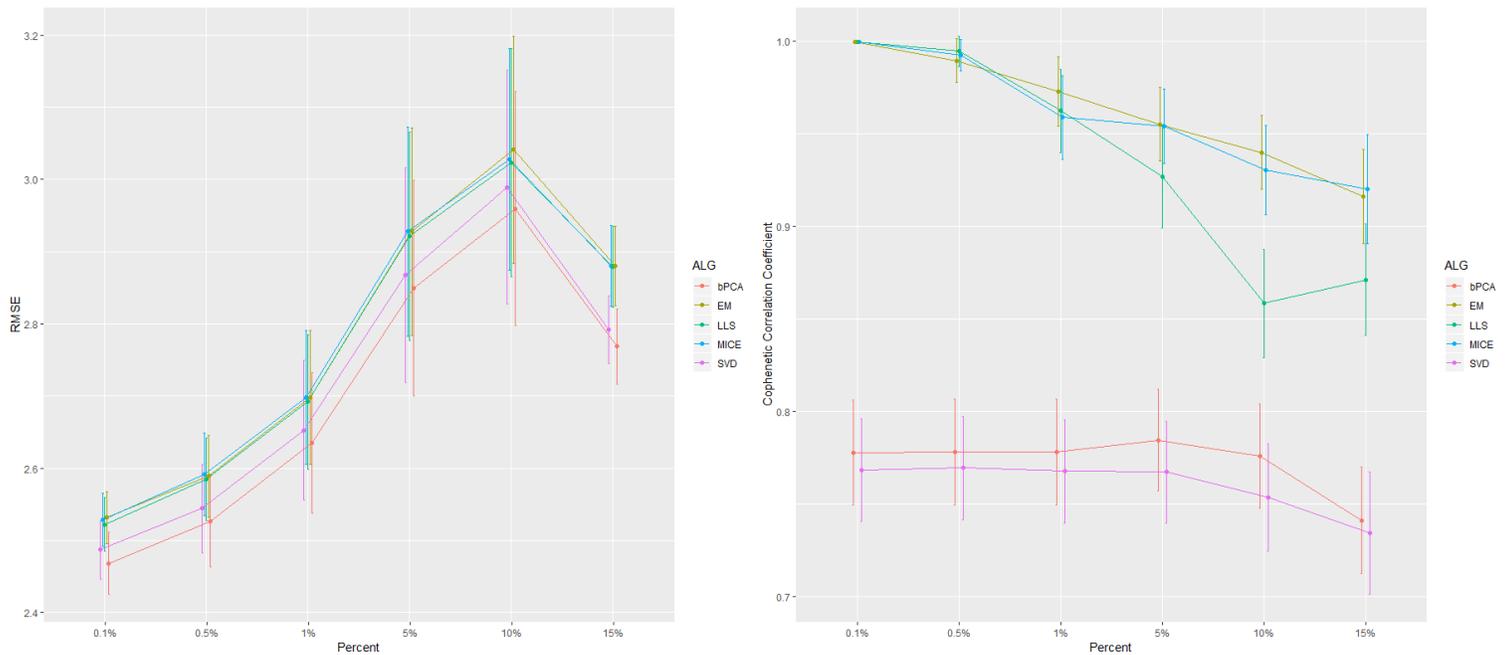

**Figure 5: Performance of Imputation Methods Across All Data Sets.** The overall capability of each imputation method over different percentages of missing values to reproduce original clustering outcomes and original values is presented. An error bar calculated from a 95% confidence interval is also depicted.

**Discussion of the Results**

The advanced imputation methods evaluated in this study are highly dependent on the type of data sets they are implemented on. Even though data sets were standardized and similar, these imputation methods also, to some degree, reflect the individual differences of the data sets tested in this study. The vast differences in solidity between the RMSE and cophenetic correlation coefficient graphs support this argument. The effect of imputation methods on clustering outcomes is great even when the imputation accuracy is high. Clusters are fundamentally different, even little changes in the wrong places can alter the final clustering outcome immensely.

**Conclusion**

Missing values often greatly affect experimental outcomes, bringing to perspective the favorable solution of imputation. As imputation accuracy grows, clustering accuracies are traditionally disregarded in assumption that preservation of representativeness precipitates preservation of clusters. However, as reflected in this study, that assumption is widely inaccurate, emphasizing the necessity of evaluating imputation method performance on their ability to reproduce natural gene groupings.

**Methods**

Each data set *m x n* (where *m* is the number of individuals and *n* is the number of genes) preprocessed through a log2 transformation and by removing rows containing missing expression values, yielding complete data sets. Further, each data set was standardized, transforming the expression values such that they form a normal distribution with a mean of 0 and standard deviation of 1.

The resulting calculations – RMS errors and cophenetic correlation coefficients – are averaged and then a confidence interval of 95% is computed to represent a range in which the true value of the missing values can be found:

$$\bar{x} \pm z \frac{s}{\sqrt{n}}$$

where $\bar{x}$ is the sample mean, $s$ is the standard deviation of the data set, $n$ is the sample size, and $z$ is the Z-value for a 95% confidence interval.

**Imputation Methods**

Five missing value estimation methods were implemented and evaluated: a method based on an expectation-maximization algorithm with a bootstrapping approach (EMB), multiple imputation by chained equations (MICE), local least squares (LLS), single value decomposition (SVD), and Bayesian Principal-Component Analysis (bPCA).

AMELIA II multiple imputation method applies an expectation-maximization algorithm with a bootstrapping approach (EMB) where imputation is iterated to generate $m$ "complete" data sets in which missing values filled with a range of estimations to demonstrate the uncertainty encountered in imputation (see [13] for more details on the EMB algorithm). After a complete data set has been produced from imputation, multiple statistical methods are available for use to combine the results. This imputation model is applied under the assumption that the complete data follows a multivariate normal distribution and the pattern of MAR or a simpler case of MCAR. [14]

The framework of multiple imputation by chained equations can be summarized in three elementary steps: imputation, analysis, and pooling. Beginning similarly to AMELIA's method, several imputed versions of one set of data is produced through multiple imputation – each data set identical in its non-missing values and unique in its missing, imputed values, reflecting the uncertainty of the estimations. The estimates are then pooled into one estimate and its variance is calculated (see [9] for details of the MICE method).

The single value decomposition imputation model uses a singular value decomposition approach, linearly combining a set of $k$ most significant mutually orthogonal expression patterns, or eigengenes, to estimate missing values. The matrix containing the eigengenes, $V^T$, is given by:

$$A_{m \times n} = U_{m \times m} \sum_{m \times n} V^T_{n \times n}$$

Through a repetition of this process, a "final product" is obtained when the total change in the matrix is below the threshold of 0.01 (see [12] for details on the SVD imputation method).

The LLS estimation method utilizes local least squares where the $k$-nearest neighbor gene vectors are found based on the $L_2$–norm by the Pearson correlation coefficient, given by:

$$r = \frac{\sum_{i=1}^{n}(x_i - \bar{x})(y_i - \bar{y})}{\sqrt{n \sum x_i^2 - (\sum x_i)^2} \sqrt{n \sum y_i^2 - (\sum y_i)^2}}$$

Then, the missing value $a$ is imputed by the optimal linear combination of the selected gene vectors found by LLS regression:

$$a = b^T x = b_1 x_1 + b_2 x_2 + \cdots + b_n x_n$$

where $b$ is a $k$-neighboring gene vector and $x_i$ are the coefficients of the linear combination. In the case where multiple missing values are present in a gene, this process is magnified (see [11] for details on the LLS imputation method).

In short, the Bayesian Principal-Component Analysis method estimates with three processes: principal component regression (PCR), Bayesian estimation, and an expectation-maximization algorithm. bPCA represents $d$-dimensional gene expression vectors $y$ as a linear combination of a set of linearly uncorrelated principal axis vectors $w$ obtained through an orthogonal transformation:

$$y = \sum_{l=1}^{K} x_l w_l + \varepsilon$$
$$(1 \leq l \leq K, and\ K < d)$$

such that $x_l$ denotes the factor scores, $\varepsilon$ denotes the residual error, and the principal axis vectors are characterized by $W = (W^{obs}, W^{miss})$, where $W^{obs}$ denotes the observed data and $W^{miss}$ denotes the missing data. For a known number $K$ in the expression vector $y$, the factors scores $x = (x_1, x_2, \ldots, x_n)$ can be calculated by minimizing the residual error:

$$err = \|y^{obs} - W^{obs}x\|^2$$

Then, the missing value in the expression vector $y$ can be estimated with:

$$y^{miss} = W^{miss}x$$

where:

$$x = (W^{obsT}W^{obs})^{-1}W^{obsT}y^{obs}$$

to the expectation with respect to the estimated posterior given by:

$$\hat{y}^{miss} = \int Y^{miss} q(Y^{miss}) dY^{miss}$$

(see [10] for more details on the bPCA method).

**Imputation Validation**

The RMSE is used to evaluate imputation accuracy. Conventionally, this value is used to measure the magnitude of difference between original and predicted values. [16] This statistic is given by:

$$RMSE = \sqrt{\frac{\sum_{t=1}^{T}(x_{1,t} - x_{2,t})^2}{T}}$$

where $x_{1,t}$ and $x_{2,t}$ are the two values that are being compared and $T$ is the number of predictions. A RMSE value closer to 0, represents a better fit of data.

**Clustering Method**

Gene expressions were clustered through agglomerative hierarchal clustering with Ward's minimum variance's clustering structure, which emerged as the strongest among the choices. [20] Ward's method is given by:

$$d_{ij} = d(\{X_i\}, \{X_j\}) = \|X_i - X_j\|^2$$

Fundamental to a bottom-up method of clustering, Ward's method minimizes the total within-cluster variances as a method of selecting pairs of clusters to merge after each step. [19] An Euclidean distance is used as a metric for this clustering method, which is given by:

$$\|a - b\|_2 = \sqrt{\sum_i (a_i - b_i)^2}$$

**Cluster Validation**

Cophenetic correlation coefficient was used to determine how rigorously a dendrogram produced from agglomerative hierarchical clustering preserves pairwise distances of the original dendrogram in the reconstructed dendrograms. [17], [18] In the original data $X_i$ that produced a dendrogram $T_i$, the cophenetic correlation coefficient $c$ is given by:

$$c = \frac{\sum_{i<j}(x(i,j) - \bar{x})(t(i,j) - \bar{t})}{\sqrt{[\sum_{i<j}(x(i,j) - \bar{x})^2][\sum_{i<j}(t(i,j) - \bar{t}^2]}}$$

A correlation coefficient closer to 1 represents more similarity between dendrograms.

**Data Sets**

This study uses four gene expression microarray data sets (see Table 1). Data set 54536 is from a recently published study that analyzes the peripheral blood of Parkinson's patients. [21] Data set 2490 is from a published study that analyzes epithelial cells of cigarette smokers. [22] Data set 5473 is from a recently published study that analyzes skeletal muscle tissue. [23] Data set 6248 is from a recently published study that analyzes liver tissue. [24]

**Acknowledgements**

The author thanks Dr. S. Piccolo for the support and inspiration and Dr. J. Guo for the continuous encouragement and motivation that largely propelled this study.